\documentclass[%
 reprint,
 amsmath,amssymb,
 aps,
]{revtex4-2}

\usepackage{graphicx}
\usepackage{dcolumn}
\usepackage{bm}

\usepackage{color}     
\usepackage{amsmath}
\usepackage{multirow}  

\begin{document}

\preprint{APS/123-QED}

\title{Effective $\beta$-decay rates of $r$-process waiting points in realistic stellar environments}

\author{Qi-Ye Hu}%
\affiliation{School of Physical Science and Technology, Southwest University, Chongqing 400715, China}%

\author{Long-Jun Wang}
\email{longjun@swu.edu.cn}
\affiliation{School of Physical Science and Technology, Southwest University, Chongqing 400715, China} 

\author{Yang Sun}
\affiliation{School of Physics and Astronomy, Shanghai Jiao Tong University, Shanghai 200240, China}%

\date{\today}

\begin{abstract}
  Reliable nuclear weak rates are key inputs for understanding the origin of heavy elements and constraining the environments of the corresponding stellar nucleosynthesis. We present the effective stellar $\beta^-$-decay rates of the $N=50, 82, 126$ $r$-process waiting-point nuclei in realistic stellar environments with high temperature, high density and strong magnetic field. Both allowed and first-forbidden transitions are considered, and transitions from the low-lying states of parent nuclei due to the thermal population are taken into account properly. The stellar $\beta^-$-decay rates of the $N=50, 82$ waiting points are not sensitive to stellar temperature, while those of the $N=126$ waiting points increase rapidly with stellar temperature. With the increase of stellar density, the electron chemical potential increases accordingly, which leads to reduction of the stellar $\beta$-decay rates. Besides, the stellar $\beta$-decay rates are found to increase rapidly with the magnetic field $B$ when $B \gtrsim 10^{14}$ G. Depending on the stellar temperature, density and magnetic field, the rates may vary by several orders of magnitude, which indicates that dynamic $\beta$-decay rates for corresponding stellar conditions may be indispensable inputs for understanding the $r$-process nucleosynthesis. 
\end{abstract}

\maketitle



The understanding of the origin of heavy elements demands interdisciplinary information and cooperation. About half of the heavy elements are produced by the rapid neutron-capture process ($r$ process), which may take place in core-collapse supernovae, neutron-star merger and/or magnetar \cite{Kajino_2019_PPNP, r_process_RMP_2021, r_process_magnetar}. These possible sites probably involve high temperature, high density and strong magnetic field, which should affect the important neutron-capture and $\beta$-decay nuclear processes during the $r$ process. 

\begin{figure}
\begin{center}
  \includegraphics[width=0.49\textwidth]{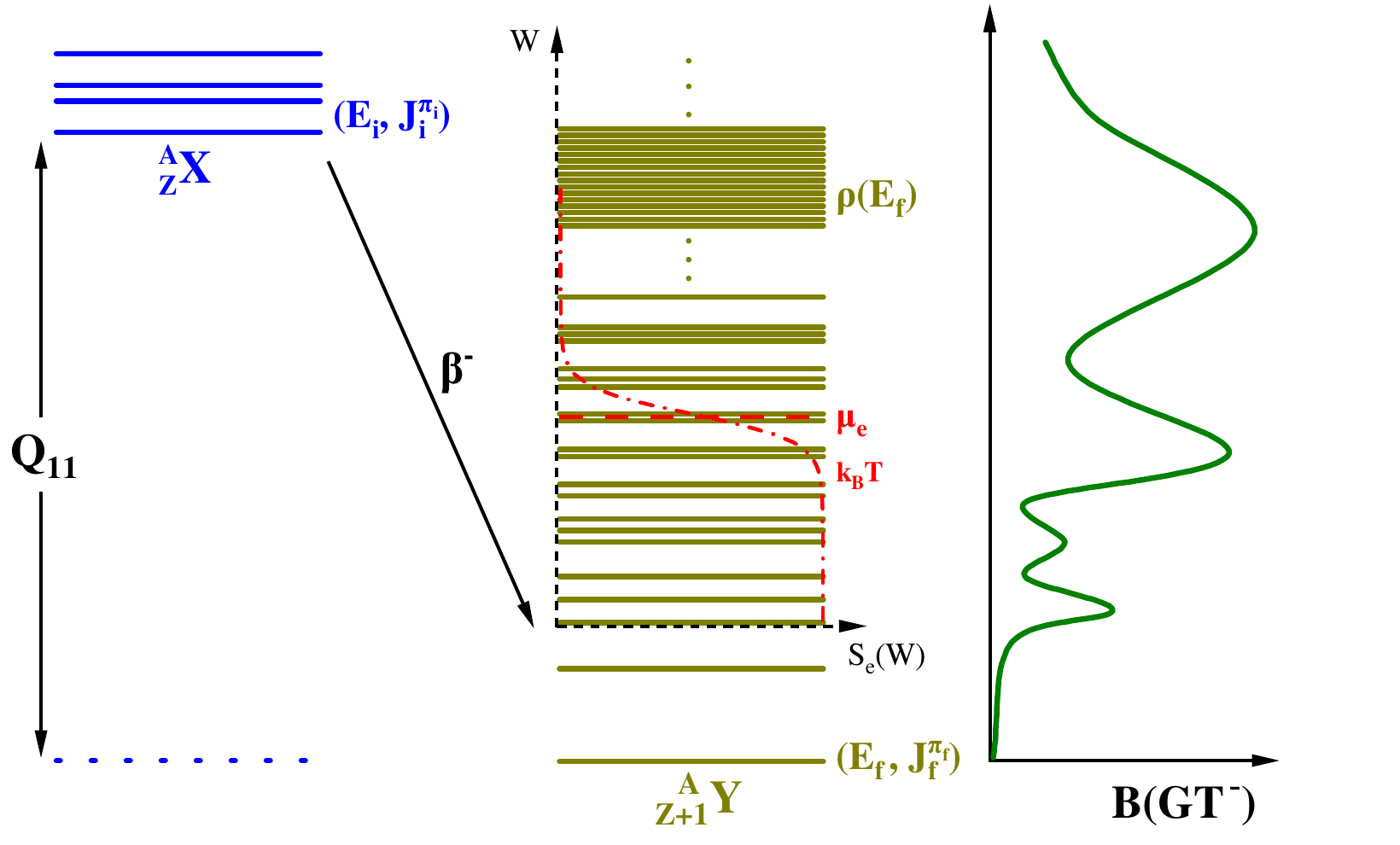}
  \caption{\label{fig:schematic} A schematic diagram for the $\beta^-$ decay scheme of some neutron-rich nucleus $^A$X at stellar conditions. The phase space of the emitted electron is determined by the $B(\text{GT})$ distribution, chemical potential $\mu_e$ and temperature $T$. The rapid increase of nuclear level density $\rho(E_f)$ with excitation energy $E_f$ is illustrated.  } 
\end{center}
\end{figure}

Reliable nuclear $\beta$-decay rates in the above stellar environments are key inputs for understanding the $r$ process. The high temperature leads to thermal population of low-lying states of parent nuclei (see Fig. \ref{fig:schematic}), the transitions from the corresponding low-lying states may enhance the $\beta$-decay rates. With the increase of stellar (electron) density, the increasing electron chemical potential would shrink the phase space of the emitted electron (see Fig. \ref{fig:schematic}) and reduce the $\beta$-decay rates. Besides, the $\beta$-decay rates are expected to increase rapidly with magnetic field \cite{Lai_ApJ_1991, Lai_RMP_2001, Zhang_Jie_2006_CPC, Famiano_2020_ApJ, Famiano_2022_ApJ}, which is crucial as for example, neutron-star mergers may have strong magnetic field with $B \approx 10^{14-18}$ G \cite{Price_2006_Science, Kiuchi_PRD_2015, Ruiz_2020_PRD}.

Along the $r$-process path the matter flow has to slow down near the neutron-rich $N=50, 82, 126$ isotones to wait for their $\beta$ decays. The stellar $\beta$-decay rates of these waiting points can not only determine the $r$-process timescale, but also be used to constrain the $r$-process site and astrophysical condition, where forbidden transitions may contribute significantly to the terrestrial rates \cite{Jon_Engel_PRC_1999_waiting_point, Suzuki_2012_PRC, Zhi_FF_PRC_2013, Roben_PRC_2024_first_forbidden}. In this Letter, we present the effective stellar $\beta^-$-decay rates of the $N=50, 82, 126$ $r$-process waiting points in stellar environments with high temperature, high density and strong magnetic field, for the first time, based on the state-of-the-art shell model with both allowed and first-forbidden transitions as well as transitions from the low-lying states considered properly.


Stellar nuclear $\beta^-$-decay rate with high temperature, high density and strong magnetic field can be obtained by,
\begin{eqnarray} \label{eq.total_lambda}
  \lambda^{\beta^-} = \sum_{if} \frac{(2J_{i}+1)e^{-E_{i}/(k_{B}T)}}{G(Z,A,T)} \lambda^{\beta^-}_{if},
\end{eqnarray}
where transitions from initial ($i$) and final ($f$) states of parent and daughter nuclei (with spin-parity assignments $J_i^{\pi_i}, J_f^{\pi_f}$ and excitation energies $E_i, E_f$) are considered, $k_{B}$ labels the Boltzmann constant and $T$ is the environment temperature. $G(Z, A, T) = \sum_i (2J_i + 1)\text{exp}(-E_{i}/(k_{B}T))$ is the partition function. In the case with magnetic fields, electron distribution is not isotropic in momentum space, electron motion in the directions perpendicular to the magnetic field is quantized into Landau levels. The individual rate can be derived as \cite{Lai_ApJ_1991, Xiao_Wang_PRC_2024},
\begin{align} \label{eq.lambda_if}
  \lambda^{\beta^-}_{if(B)}  =& \frac{\ln 2}{K} \frac{B^\ast}{2} \sum_{n=0}^{N_{\text{max}}} (2-\delta_{n0}) \int_{0}^{ p_{znu} } C(W_n) (Q_{if}-W_n)^2 \nonumber \\
                              & \qquad \quad \times F_0(Z+1, W_n) (1 - S_e(W_n)) dp_z, 
\end{align}
where the constant $K=6144 \pm 2$ s \cite{Hardy_2009_PRC} is adopted, and $B^\ast \equiv B/B_c$ is the dimensionless magnetic field strength in unit of the critical magnetic field $B_c = m_e^2 c^3 / e \hbar \approx 4.414 \times 10^{13}$ G \cite{Lai_ApJ_1991}. $n$ labels the $n$-th Landau level with the spin degeneracy $(2-\delta_{n0})$ \cite{Lai_ApJ_1991, Lai_RMP_2001} which runs over from $0$ to the maximum $N_{\text{max}} = (Q_{if}^2 - 1) / 2B^\ast$, where $Q_{if} = (M_p - M_d + E_i -E_f ) / m_e c^2 $ is the (dimensionless) available total energy for leptons in individual transition, with $M_p (M_d)$ being the nuclear mass of parent (daughter) nucleus. The integral variable $p_z$ is the momentum component in the direction of the magnetic field and the integral limit $p_{znu} = \sqrt{ Q^2_{if} - 1 - 2nB^\ast }$. $W_n = \sqrt{p^2_z + 1 + 2nB^\ast}$ is the electron energy and $F_0$ is the Fermi function that accounts for the Coulomb distortion of the electron wave function near the nucleus \cite{Fuller1980, Fermi_func_1983}.

In Eq. (\ref{eq.lambda_if}), the electron and positron distribution functions follow the Fermi-Dirac distribution as,
\begin{eqnarray} \label{eq.Se}
  S_{e/p} (W_n) = \frac{1}{\text{exp}[(W_n \mp \mu_e) / k_B T] + 1} ,
\end{eqnarray}
where the electron chemical potential $\mu_{e}$ in the case with magnetic field is determined as follows,
\begin{eqnarray} \label{eq.mue}
  \rho Y_e = \frac{B^\ast}{2\pi^2 N_A \lambdabar_e^3} \sum_{n=0}^{N_{ \text{lim} }} (2-\delta_{n0}) \int_{0}^{\infty} (S_e - S_p) dp_z .
\end{eqnarray}
where $\rho Y_e$ labels the electron density, $N_A$ represents the Avogadro's number and $\lambdabar_e$ the reduced Compton wavelength of electron. $N_{\text{lim}} \approx \mu_e^2 / 2B^\ast$ is the effective limit of the number of Laudau levels. 

The $C(W_n)$ in Eq. (\ref{eq.lambda_if}) is the shape factor of individual nuclear transitions including both allowed Gamow-Teller (GT) and first-forbidden transitions, which is the key nuclear-structure input \cite{Xiao_Wang_PRC_2024}. For allowed GT transitions with selection rule $|J_i - J_f| = 0, 1$ and $\pi_i \pi_f = +1$, $C(W_n)$ does not depend on the electron energy and is equal to the reduced GT strength $B(\text{GT}^-)_{if}$, which is usually evaluated by nuclear transition matrix element of the GT operator. For first-forbidden transitions with selection rule $|J_i - J_f| = 0, 1, 2$ and $\pi_i \pi_f = -1$, $C(W_n)$ has explicit energy dependence which is determined by nine different nuclear transition matrix elements. Reliable calculations and predictions of the shape factors $C(W_n)$ requires accurate nuclear wave functions in the laboratory frame with good spin and parity, large model and configuration spaces, as well as the consideration of possible high-order terms of the transition operators \cite{LJWang_current_2018_Rapid, Gysbers_2019_Nat_Phys}, to account for the quenching problems of nuclear matrix elements in both allowed and first-forbidden transitions. In this work, the shape factors are calculated by the projected shell model (PSM) that has both large model space and large configuration space \cite{LJWang_2014_PRC_Rapid, LJWang_2016_PRC}, good spin and parity are obtained by exact projection techniques, and both allowed and first-forbidden transitions are considered \cite{LJWang_2018_PRC_GT, LJWang_PLB_2020_ec, LJWang_2021_PRL, LJWang_2021_PRC_93Nb, BLWang_1stF_2024, ZRChen_PLB2024, Hu_2024_arXiv_rp_magnetic}. The calculations details are shown in the Supplemental Material \cite{Suppl_Material}.


\begin{figure*}
\begin{center}
  \includegraphics[width=1.00\textwidth]{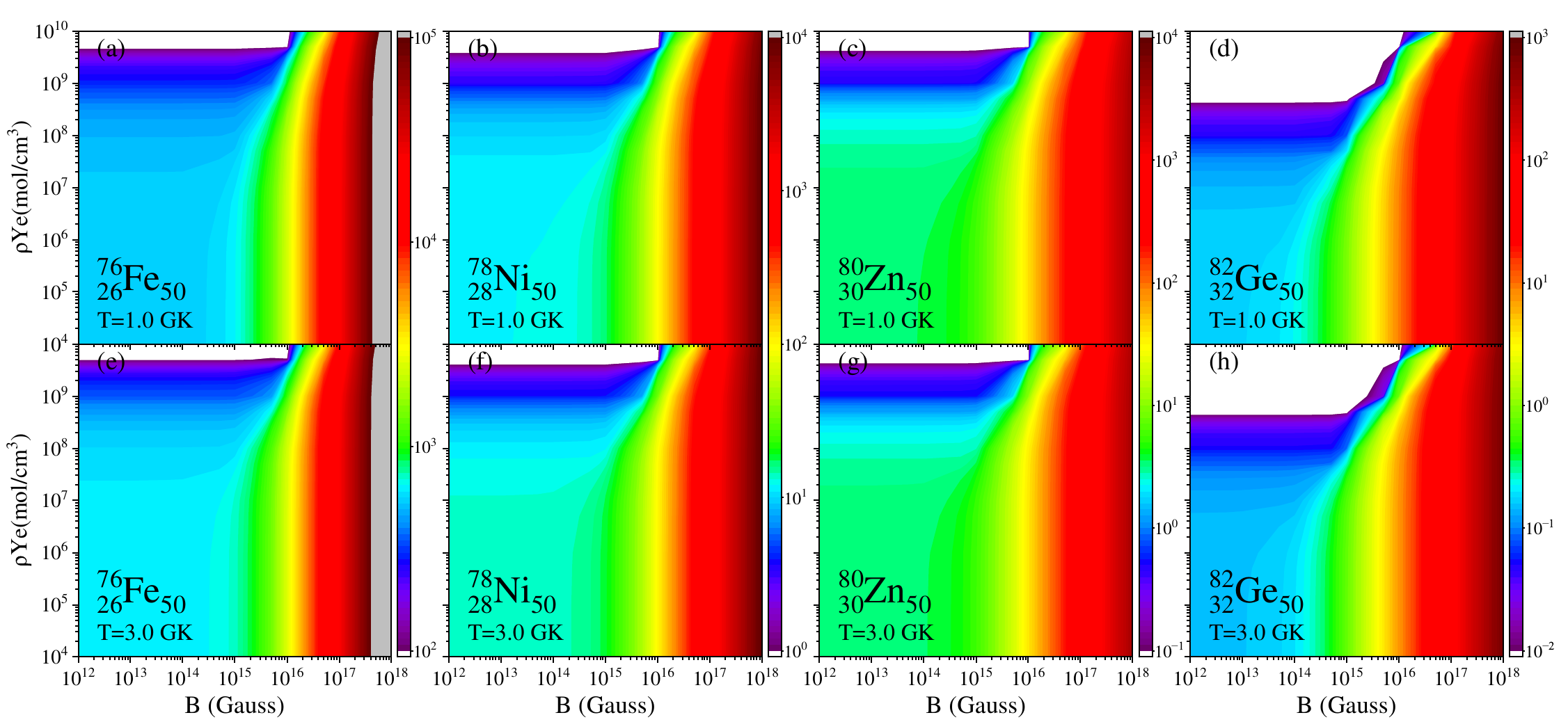}
  \caption{\label{fig:N50} The effective $\beta^-$-decay rates of the $N=50$ $r$-process waiting points at stellar temperature $T=1.0, 3.0$ GK as a function of the possible stellar electron density $\rho Y_e$ (in mol/cm$^3$) and magnetic field $B$ (in Gauss). See the text for details.  }
\end{center}
\end{figure*}

\begin{figure*}
\begin{center}
  \includegraphics[width=1.00\textwidth]{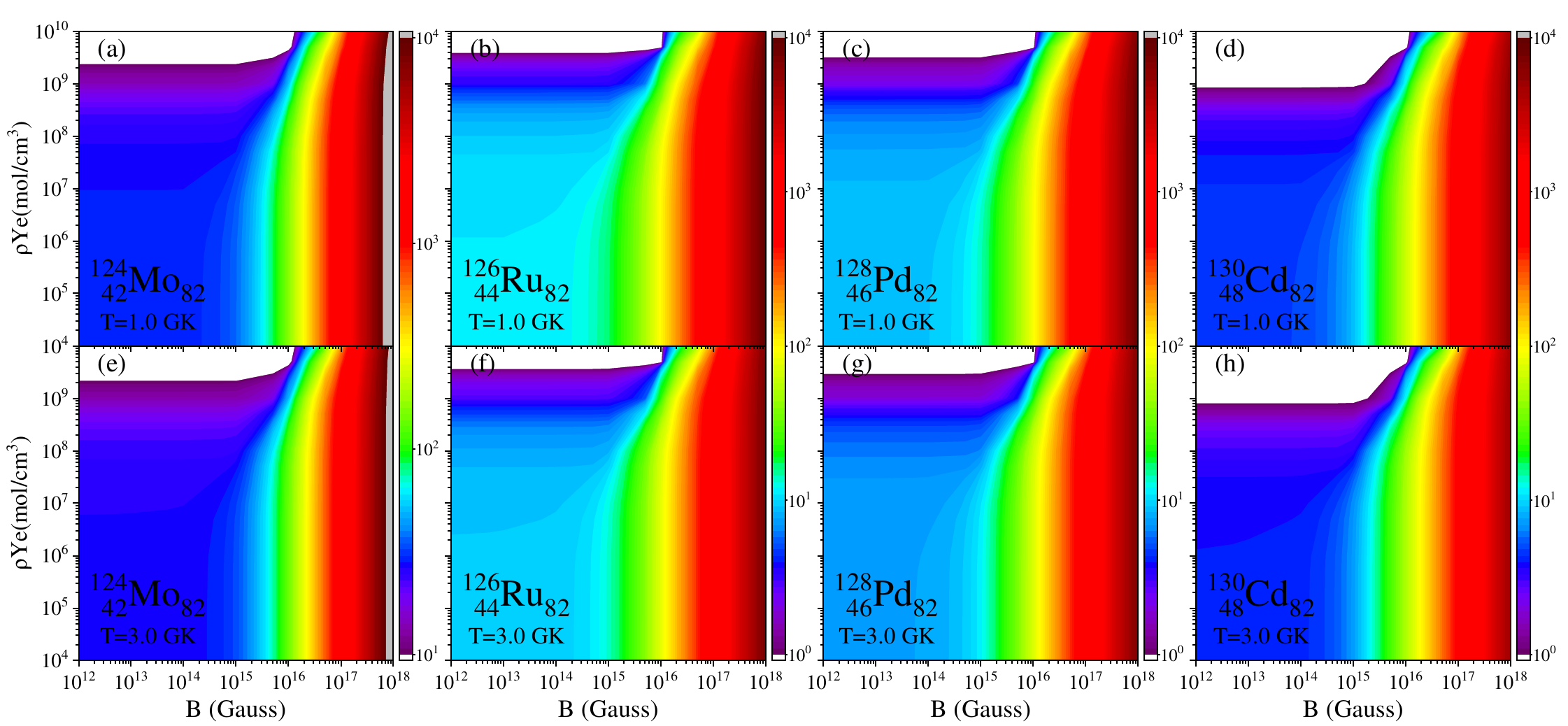}
  \caption{\label{fig:N82} The same as Fig. \ref{fig:N50} but for the $N=82$ $r$-process waiting points. } 
\end{center}
\end{figure*}

\begin{figure*}
\begin{center}
  \includegraphics[width=1.00\textwidth]{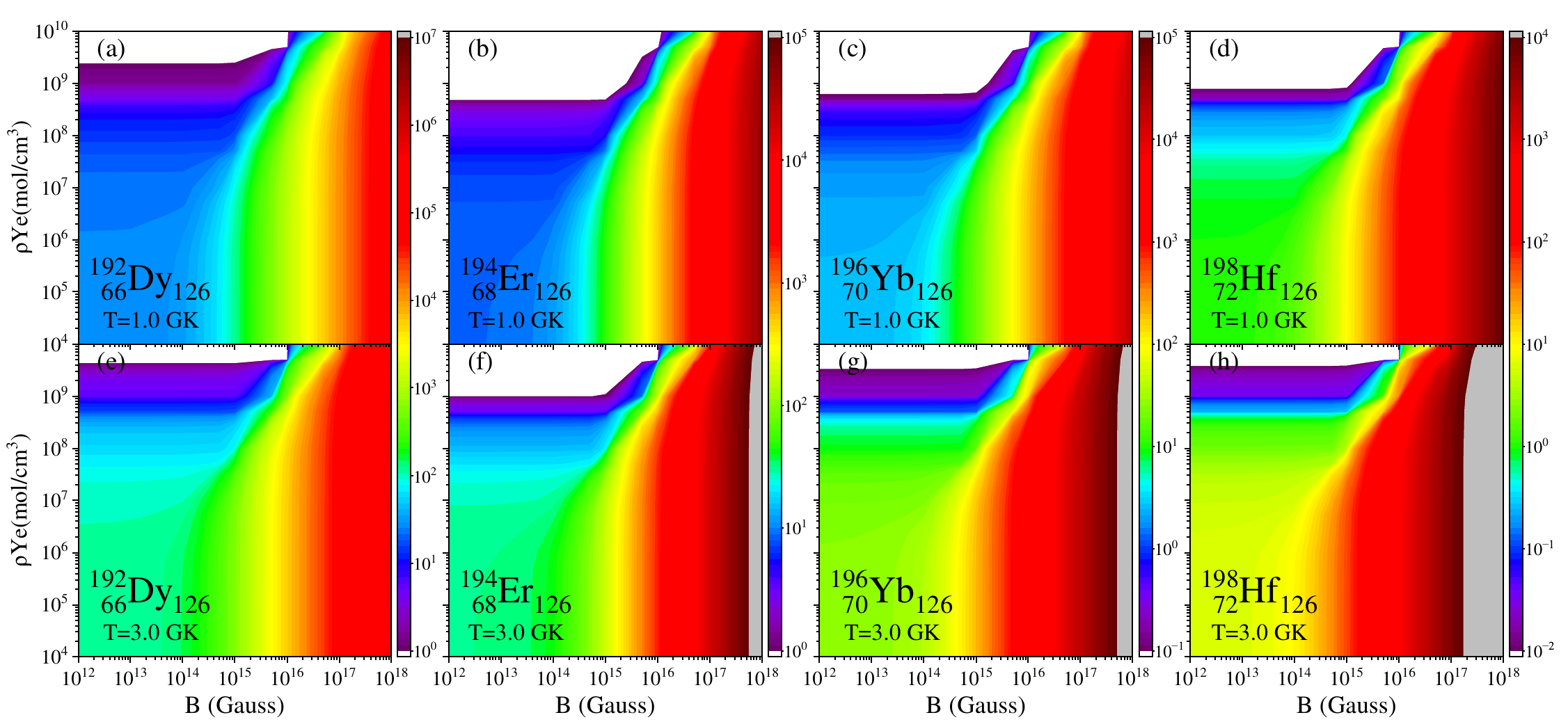}
  \caption{\label{fig:N126} The same as Fig. \ref{fig:N50} but for the $N=126$ $r$-process waiting points. }
\end{center}
\end{figure*}

Before discussions of the stellar $\beta^-$-decay rates, it is worth mentioning that with increasing magnetic field $B$, the chemical potential $\mu_e$ would remain constant at weak $B$, and then decrease when $B \gtrsim 10^{14-15}$ B \cite{Hu_2024_arXiv_rp_magnetic, Suppl_Material}. The reason is as follows. For weak $B$ cases, for specific density $\rho Y_e$ and temperature $T$, the degeneracy of state density for each Landau level or a cylinder (which is proportional to $B^\ast$) increases rapidly with increasing $B$, while more and more Landau levels are pushed outside the Fermi sphere, and the effective number of Landau levels within the Fermi sphere (i.e. $N_{\text{lim}} + 1$) decreases rapidly. The balance of these two aspects keeps the Fermi sphere, $\mu_e$, unchanged. For strong $B$ cases, with increasing $B$, the degeneracy of each Landau level increases in the same way, while the effective number of Landau levels within the Fermi sphere becomes too low and cannot reduce in the same way to balance the former, so that the Fermi sphere (i.e., $\mu_e$) has to shrink to keep the density $\rho Y_e$ on the left hand side of Eq. (\ref{eq.mue}) invariable.

Figures \ref{fig:N50}-\ref{fig:N126} illustrate the effective $\beta^-$-decay rates of the $N=50, 82, 126$ $r$-process waiting points at possible stellar temperature $T$, electron density $\rho Y_e$ and magnetic field $B$, from our PSM calculations. For low $\rho Y_e$, zero $T$ and $B$, the data of terrestrial half-lives of $^{82}$Ge and $^{130}$Cd can be well reproduced by our PSM calculations, and those of $^{78}$Ni, $^{128}$Pd and $^{80}$Zn can be described within a factor of two or three. Although the first-forbidden transitions are predicted to reduce the terrestrial half-lives of $N=82$ and/or $N=126$ waiting points by several times by the shell model in Ref. \cite{Suzuki_2012_PRC} and the quasiparticle random phase approximation (QRPA) in Ref. \cite{Roben_PRC_2024_first_forbidden}. Our PSM calculations predict that the first-forbidden transitions only contribute $\lesssim 30\%$ of the total terrestrial rates, which is similar to the shell model calculations in Ref. \cite{Zhi_FF_PRC_2013}. This indicates that the allowed GT transitions dominate and Fig. \ref{fig:schematic} is helpful for understanding the stellar rates. 

At stellar conditions, with increasing $T$ (e.g., from 1 to 3 GK), the stellar $\beta^-$-decay rates of the $N=50, 82$ waiting points show nearly no change as seen from Figs. \ref{fig:N50} and \ref{fig:N82}. The reason is that although the low-lying states of the $N=50, 82$ waiting points may be thermally populated with increasing $T$, their $B(\text{GT}^-)$ distributions are very similar to those of the ground states \cite{Suppl_Material}. However, the stellar rates of the $N=126$ waiting points increase rapidly with $T$ as seen from Fig. \ref{fig:N126}. When $T$ increases from 1 to 3 GK, the stellar rates of $^{192}$Dy, $^{194}$Er and $^{198}$Hf increase by about five times, and those of $^{196}$Yb increase by one order of magnitude. The reasons are twofold, on one hand, the $J_i^{\pi_i}$ of the first excited states is predicted to be $2^+$, for which the degeneracy $(2J_i+1)$ is five times than that of the $0^+$ ground states. This indicates, as seen from Eq. (\ref{eq.total_lambda}), these first excited states can be thermally populated effectively with increasing $T$. On the other hand, the $J_i^{\pi_i}=2^+$ excited states can connect much more final states by the GT operator than the case of ground states, and the $B(\text{GT}^-)$ distributions from these excited states are predicted to be larger than those from the ground states by more than one order of magnitude \cite{Suppl_Material}. 

For the variation of stellar density, with increasing $\rho Y_e$, in the cases of weak $B$ the chemical potential $\mu_e$ does not change with increasing $B$, as mentioned above, while $\mu_e$ will increase rapidly with increasing $\rho Y_e$. This indicates that the phase space of the emitted electron would shrink rapidly with $\rho Y_e$, as seen from Eq. (\ref{eq.lambda_if}) and Fig. \ref{fig:schematic}, leading to rapid decrease of stellar $\beta^-$-decay rates with $\rho Y_e$, which can be seen from the left sides in each panels of Figs. \ref{fig:N50}-\ref{fig:N126}. Specifically, as seen in Fig. \ref{fig:N50}(c), when $\rho Y_e$ increases from $10^7$ to $10^{10}$ mol/cm$^3$, the rate of $^{80}$Zn decreases from 10/s to $\lesssim 0.1$/s.  

Now we discuss the change of stellar $\beta^-$-decay rates with the increasing magnetic field. For the case of low $\rho Y_e$, the $\mu_e$ is very low so that the decrease of $\mu_e$ with increasing $B$ will not affect the phase space of emitted electron. In this case, the $\beta^-$-decay rates are found to remain unchanged and then begin to increase suddenly and rapidly at some pint $B_\text{onset}$. The reason is similar to that of $\beta^+$-decay rates of the $rp$-process waiting points \cite{Hu_2024_arXiv_rp_magnetic}. With rapidly increasing $B$, the degeneracy of state density for each Landau level (which is proportional to $B^\ast$) increase rapidly with $B$, while the maximum number of Landau levels within the phase space (determined by $Q_{\text{max}}$ corresponds to the maximum $Q_{if}$ when $B(\text{GT}^-)_{if}$ is not negligible), i.e., $N_{\text{max}} + 1 = (Q_{\text{max}}^2 - 1) / 2B^\ast + 1$, decreases rapidly in the same way, as seen from Eq. (\ref{eq.lambda_if}). The balance of the two keeps the stellar $\beta^-$-decay rates unchanged. When $B$ is strong enough so that $N_{\text{max}} + 1 $ is reduced to a small number (for example, $\approx 5$), the balance is broken and the rates begin to increase with $B$ proportionally, from which one can estimate that $B_\text{onset} \approx Q_{11}^2 \times 4.4 \times 10^{12}$ G. For $^{128}$Pd and $^{196}$Yb with $Q_{\text{max}} = 16$ and $10$ \cite{Suppl_Material}, one gets $B_\text{onset} \approx 10^{15}$ G and $\approx 4 \times 10^{14}$ G, which are consistent with Fig. \ref{fig:N82} (c) and Fig. \ref{fig:N126} (c).

For the case of high $\rho Y_e$ with large $\mu_e$, at weak $B$, the phase space of emitted electron is suppressed and the $\beta^-$-decay rates are small. With the increasing $B$, on one hand, the $\mu_e$ decreases rapidly, releasing larger and larger phase space which tends to increase the $\beta^-$-decay rates. On the other hand, when $B > B_\text{onset} $ the effective number of Landau levels cannot reduce rapidly further while the degeneracy of state density for each Landau level still increases rapidly with $B^\ast$, which tends to increase the $\beta^-$-decay rates as well. The two aspects lead to extremely rapid increase of the $\beta^-$-decay rates. For example, as seen from Fig. \ref{fig:N126} (d), for $\rho Y_e = 10^9$ mol/cm$^3$ and $T=1$ GK, when $B$ increase from $10^{15}$G to $10^{17}$G, the $\beta^-$-decay rates of $^{198}$Hf increase rapidly extremely from $\lesssim 10^{-2}$/s to $\approx 10^{3}$/s, by about five orders of magnitude. 

As can be seen from Figs. \ref{fig:N50}-\ref{fig:N126} and the discussions that, depending on the possible stellar temperature, density and magnetic field, the $\beta^-$-decay rates of the $N=50, 82, 126$ waiting points may vary by three to seven orders of magnitude. During stellar nuclear $\beta^-$ decays, neutrinos are emitted, which are unhindered to escape and carry away energy in most cases, cooling the environments efficiently. Therefore, it is expected that the neutrino energy-loss rates of the $N=50, 82, 126$ $r$-process waiting points should vary by orders of magnitude in the similar way \cite{Suppl_Material}. This indicates that the environmental conditions and the stellar nuclear $\beta$ decays (neutrino-cooling processes) during the $r$ process will affect each other. Dynamic $\beta$-decay rates and neutrino energy-loss rates with corresponding stellar conditions may be indispensable inputs for understanding the $r$-process nucleosynthesis. If the $r$-process abundance distribution (especially the $A \approx 195$ peak) and the corresponding neutron-capture rates were known precisely, our work can be used to constrain the possible $r$-process sites and astrophysical conditions.


In summary, the $\beta$-decay rates of $r$-process waiting points can not only determine the $r$-process timescale, but also be used to constrain the $r$-process site and astrophysical condition.  We provide, for the first time, the effective stellar $\beta^-$-decay rates of the $N=50, 82, 126$ $r$-process waiting points in realistic stellar environments based on the projected shell model. Both allowed and first-forbidden transitions are considered, and the effects of possible high temperature, high density and strong magnetic field are all treated properly. The stellar rates of the $N=126$ waiting points are sensitive to stellar temperature, and the rates of all the $N=50, 82, 126$ waiting points are very sensitive to stellar density and magnetic field. The corresponding neutrino energy-loss rates show similar sensitivity as well. These rates may vary by up to seven orders of magnitude depending on the possible astrophysical conditions. Our work should be very helpful for better understanding the $r$-process sites and environments. 

\begin{acknowledgments}
  We thank J. M. Dong for motivating us to study the effects of magnetic field on stellar nuclear weak-interaction processes. This work is supported by the National Natural Science Foundation of China (Grant Nos. 12275225, 12235003). 
\end{acknowledgments}






%

\end{document}